\documentclass[useAMS,usenatbib]{mn2e} 
\usepackage{txfonts,graphicx} 
\usepackage{color} 
\usepackage{natbib,longtable,url} 
\include{definitions} 

\title[1D NLTE spectroscopic stellar parameters]{NLTE line formation of Fe in late-type
stars II: 1D spectroscopic stellar parameters} 
\author[K. Lind]{K. Lind \thanks{E-mail: klind@mpa-garching.mpg.de}$^1$, M. Bergemann$^1$, M. Asplund$^{1,2}$ \\ \\
$^1$ Max-Planck Institute for Astrophysics, Karl-Schwarzschild Str. 1, 85741, Garching, Germany \\ 
$^2$ Research School of Astronomy \& Astrophysics, Mount Stromlo Observatory, Cotter Road, Weston Creek, ACT 2611\\
}

\begin{document}

\date{Accepted Date. Received Date 03.02.2012; in original Date}

\pagerange{\pageref{firstpage}--\pageref{lastpage}} \pubyear{2011}

\maketitle

\label{firstpage}

\begin{abstract} {We investigate departures from local thermodynamic equilibrium (NLTE) in the line formation of neutral and singly ionised iron lines and their impact on  spectroscopic stellar parameters. The calculations are performed for an extensive grid of 1D MARCS models of metal-rich and metal-poor late-type dwarfs and giants. We find that iron abundances derived from Fe\,I lines are increasingly underestimated in hotter, lower surface-gravity, and more metal-poor stars, in a simple and well-defined pattern, while LTE is usually a realistic approximation for Fe\,II lines. For the vast majority of dwarfs and giants, the perturbed ionisation balance of Fe\,I and Fe\,II is the main relevant NLTE effect to consider in the determination of spectroscopic parameters, while for extremely metal-poor stars and hot giant stars significant impact is seen also on the excitation balance and on the microturbulence determination from Fe\,I lines.} \end{abstract}

%
%
\section{Introduction} 

Measuring the iron content of late-type stars is one of the primary
challenges for the field of Galactic archaeology. Due to
its large opacity contribution in cool stellar atmospheres, Fe has
come to serve as a fundamental reference point for all chemical analysis and its
interpretations. 
Furthermore, the visibility of a wealth of spectral lines with a range
of atomic properties and line strengths enables the determination of
spectroscopic stellar parameters through the excitation and
ionisation equilibria. With the knowledge of the star's iron content, effective temperature, and surface gravity, we can then acquire further information about the evolution of the star itself, i.e. determine its mass and age, and the evolution of the stellar population it resides in. 

However, the low fraction of neutral iron in stellar atmospheres makes its line formation sensitive to
departures from local thermodynamic equilibrium (LTE). Traditional LTE
analyses of FeI lines therefore tend to underestimate the true Fe
abundance, as is well described in the literature \citep[e.g.][]{Thevenin99,Gehren01,Collet05,Mashonkina11a}.  

The success of spectroscopic methods clearly depends on the realism of the atmospheric structure. Even if high precision may be obtained with a very simplistic model, accuracy may or may not. Traditional methods rely on model atmospheres calculated under the assumptions of hydrostatic equilibrium and a one-dimensional (1D) geometry. The most critical approximation is the mixing-length description of the convective energy transport. The advent of 3D, radiation-hydrodynamical simulations has shed light on the systematic uncertainties introduced by these simplifying assumptions. In particular, metal-poor stars have been demonstrated to have much cooler line-forming layers than previously thought \citep[see e.g.][and references therein]{Asplund05}.

In Paper I in this series \citep{Bergemann12} we demonstrated that
a modelling technique allowing for departures from LTE can be used to
accurately predict iron abundances and spectroscopic stellar
parameters for a set of benchmark late-type stars. 
We specifically illustrated how 
traditional 1D hydrostatic models successfully meet the ionisation and
excitation equilibria for solar metallicity stars, but evidently fail to
achieve excitation equilibrium at realistic temperatures in metal-poor stars. In particular, low excitation lines are sensitive to the atmospheric structure, and require a combination of 3D and NLTE analysis. We further
illustrated how the ionisation balance of Fe\,I and Fe\,II lines can be exploited to infer realistic spectroscopic surface
gravities and metallicities also with traditional 1D models, once the effective 
temperature has been determined by other means. Alternatively, the ionisation balance can be
used to constrain the effective temperature if the gravity is known by other means.

Due to the increasing numerical complexity, compared with the LTE case, NLTE investigations have
previously been limited to individual stars and usually only a handful
of spectral lines. For the first time, we present an extensive grid of 
calculations, with individual NLTE abundance corrections given for thousands of lines in the ultra-violet,
optical and near-infrared part of the spectrum. The calculations span
from ultra metal-poor to super-solar metallicities and include dwarfs,
subgiants, and red giant stars. The calculations can easily be extended to an 
analogous grid of spatially and temporally averaged 3D models (hereafter \textless3D\textgreater\,), once such becomes available from e.g. the \textsc{Stagger} \citep{Collet11} or \textsc{CO5BOLD} collaborations \citep{Ludwig09}. 

%
%

\begin{figure} 
\begin{center} 
\includegraphics[angle=90,scale=0.35,viewport=0.5cm 2cm 21cm 24cm]{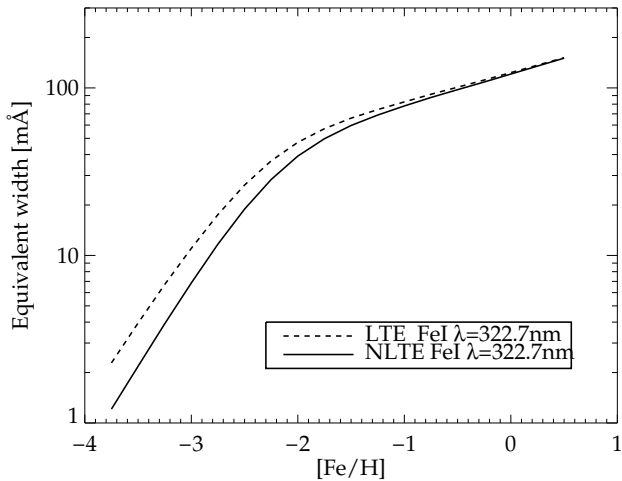}
\caption[]{Example LTE and NLTE curves-of-growth for a UV Fe\,I transition. The model parameters are $T_{\rm eff}=6500$\,K, $\rm log\it(g)\rm=4.0$, and $\xi_{\rm t}=2$\,km/s. The NLTE abundance correction at a given equivalent width, $\Delta A\rm(Fe)_{I}$, is defined as the difference between the two curves. } 
\label{Cog} 
\end{center} 
\end{figure}

\section{Methods}{\label{sec:methods}}
As is customary for late-type stars, we solve the restricted NLTE
problem given the assumption of iron as a trace
element. Some justification for this methodology is presented in
Sect.\,\ref{sec:trace}. The 1D, NLTE code \textsc{MULTI}, version 2.3, \citep{Carlsson86,Carlsson92} was
used to simultaneously solve the equations of radiative transfer and
statistical equilibrium for a model atom of neutral and singly ionised
iron.    

\begin{figure*} 
\begin{center} 
\includegraphics[scale=0.85,viewport=0cm 0cm 22cm 19cm,clip]{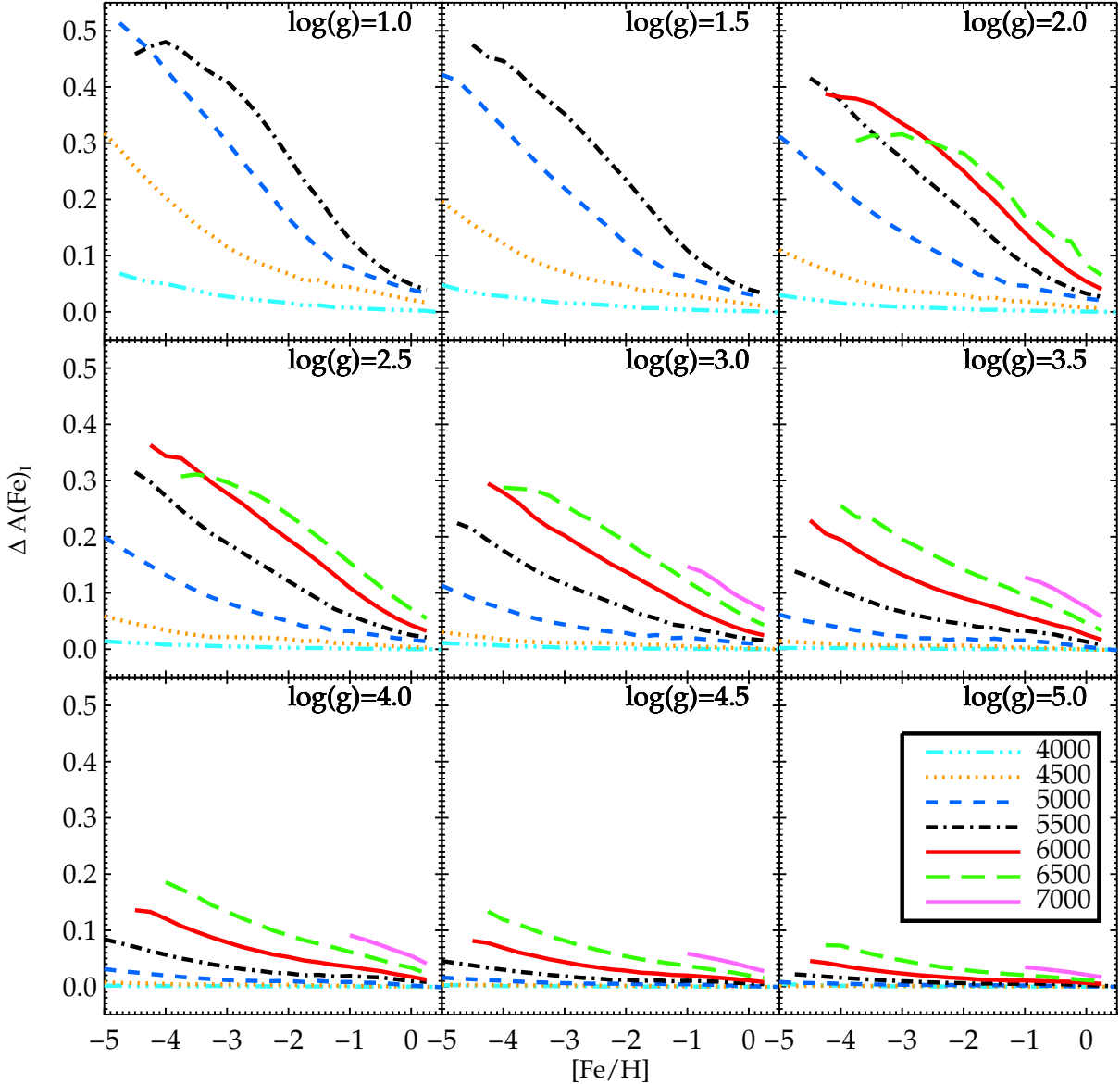}
\caption[]{The lines mark the typical sizes of NLTE effects for high-excitation ($E_{\rm exc}>2.5\rm\,eV$), unsaturated ($W_\lambda<50\rm\,m\AA\,$) Fe\,I lines and their dependence on stellar parameters. All models shown have $\xi_{\rm t}=2.0$\,km/s. } 
\label{delta} 
\end{center} 
\end{figure*}

\subsection{Model atom}{\label{sec:atom}}

The model atom used to establish the statistical equilibrium was
presented in detail in Paper I and we give only a summary here. The
atom consists of 296 levels of Fe\,I, 112 levels of Fe\,II and the
Fe\,III ground state, with experimentally determined and/or
theoretically predicted level energies. Levels close in energy are
collapsed into superlevels to save computing time. Radiative
bound-bound transition probabilities are collected from NIST\footnote{\url{http://physics.nist.gov/asd}} \citep{Ralchenko12}
and R.\,L.\,Kurucz\footnote{\url{http://kurucz.harvard.edu/}}  data bases and photo-ionisation cross-sections were
provided by M.\, Bautista (\citeyear[][and priv. communication]{Bautista97}). 
Rate coefficients for electron impact
excitation and ionisation were computed using semi-empirical
recipes \citep{vanRegemorter62,Seaton62a,Allen73,Takeda94}. 

The efficiency of excitation and ionisation (and charge exchange
reactions) induced by collisions with neutral hydrogen is commonly the
largest source of uncertainty in statistical equilibrium calculations
for late-type stars \citep[see discussion in e.g.][]{Asplund05}. In the absence of better alternatives, we adopt
the Drawin formulae \citep{Steenbock84} for these processes, and a free scaling parameter
($\rm S_H$) that sets the absolute efficiency of the collisions. As
explained in Paper I, the scaling factor can be
constrained when confronted with observations of stars with surface gravities
and effective temperatures accurately
known through independent methods. We discuss the impact of our choice of
scaling factor, $\rm S_H=1.0$, in Sect.\,\ref{sec:error}. 

When the atomic level populations had reached convergence, high-resolution NLTE
line profiles of 3239 lines of Fe\,I and 107 lines of Fe\,II were
computed. The lines were selected based on having a non-negligible
line strength at solar metallicites and all fall in the wavelength range
$3200-9300\AA\,$. We impose no further constraints based on the accuracy
of atomic data or contribution of blending species in stellar
spectra. Hence, lines may or may not be suitable for abundance
analysis in a given star. The oscillator strengths are adopted from
\citep{Kurucz07}. For 1479 Fe\,I lines and 95 Fe\,II lines, ABO theory and the quantum mechanical treatment of pressure broadening by H collisions is
implemented for broadening by neutral hydrogen \citep{Barklem00b,Barklem05b}. In the absence of these
data, we resort to \citet{Unsold55} theory with an enhancement factor of 1.5.
Stark broadening is neglected. 

\subsection{Model atmospheres}{\label{sec:atmosphere}}

We use a grid of 1D, hydrostatic model atmospheres computed with the
MARCS code \citep{Gustafsson08}. The models all have standard
composition\footnote{See \url{marcs.astro.uu.se}.}, i.e. scaled solar \citep{Grevesse07} with gradual alpha-enhancement up to 0.4\,dex at
$\rm[Fe/H]\leq-1.0$. The grid dimensions are specified in Table
\ref{tab:grid}. Models that are missing from the original grid are obtained by
interpolation or, in some cases, substitution\footnote{When needed and
justified we do not enforce consistency between the microturbulence used to
compute the model atmosphere and that used to perform the NLTE
calculations, i.e. missing models are replaced with ones of higher/lower microturbulence velocities.}.   

\begin{table}
      \caption{Dimensions of model atmosphere grid}
         \label{tab:grid}
         \centering
         \begin{tabular}{llll}
                \hline\hline
                Parameter           &   Min.  & Max. & Step  \\
                                \hline
                $T_{\rm eff}$[K]    & $4000$  &  $8000$    & $500$   \\
                $\log{g}^{\rm(a)}$    [cgs]  & $1.0$   &  $5.0$     & $0.5$   \\
                $\rm[Fe/H]$         & $-5.0$  &  $0.5$     & $0.25$ \\
                $\xi_{\rm t}$ [km/s]& $1.0$   &  $2.0$     & $1.0$ 
   \\
                   \hline
                \multicolumn{4}{l}{$\rm^{(a)}$ $\log(g)\ge 2$ for $T_{\rm
eff}\ge6000$\,K}         \\
                \multicolumn{4}{l}{\ \ \ \ \  $\log(g)\ge 3$ for $T_{\rm
eff}\ge7000$\,K}       \\
                \multicolumn{4}{l}{\ \ \ \ \  $\log(g)\ge 4$ for $T_{\rm
eff}=8000$\,K}       \\
         \end{tabular}
\end{table}

\subsection{NLTE abundances}{\label{sec:abund}}

Here we base the comparison between LTE and NLTE on equivalent
widths ($W_\lambda$), rather than synthetic line profiles. Note however that differences in lines profiles become important
 for saturated lines observed at high resolution. For
each grid point, we compute the LTE and NLTE
equivalent widths for the iron abundance that is identical to the one
adopted for the atmosphere. The line strengths increase smoothly with
increasing metallicity in a manner similar to a traditional
curve-of-growth. Fig. \ref{Cog} illustrates the curves derived for a 
spectral line at a certain effective temperature, surface gravity, and
microturbulence.   
The NLTE abundance correction is defined as the difference
between the NLTE and LTE curves at a given equivalent width. For practical
reasons we introduce a cut towards weak lines at 1\,m\AA\,. Lines weaker than this  are not
discussed. Our equivalent width integrations are numerically reliable up to $\sim$1000\,m\AA\,.

Inspection of the NLTE abundance corrections determined for a given
set of stellar parameters reveals a strong similarity between lines of
 similar excitation potential and line strength. In the
following discussion we will use the concept of a mean NLTE abundance correction, restricting ourselves to all lines below a certain limit in
equivalent width (here $50$\,m\AA\,) and above a certain limit in excitation potential (here $2.5$\,eV) of
the lower level. The first restriction is made because 
abundances of strong lines are through common practice forced into agreement with weaker lines
by fine-tuning of the microturbulent velocity. The absolute abundances and NLTE
corrections derived from microturbulence-sensitive lines are thus of
lesser importance. The second restriction to high-excitation lines is
motivated by their lower sensitivity to 3D effects compared to
low-excitation lines, as discussed in Paper I. However, in 1D, our atomic model predicts only minor differential NLTE effects for lines of high and low excitation potential. Figure \ref{delta}
illustrates how the mean NLTE abundance correction of un-saturated, high-excitation Fe\,I lines ($\Delta\rm A(Fe)_I$) varies with stellar parameters.  

\section{Discussion}{\label{sec:discussion}} 

\begin{figure*} 
\begin{center} 
\includegraphics[angle=90,scale=0.7,viewport=0cm 1cm 15cm 25cm]{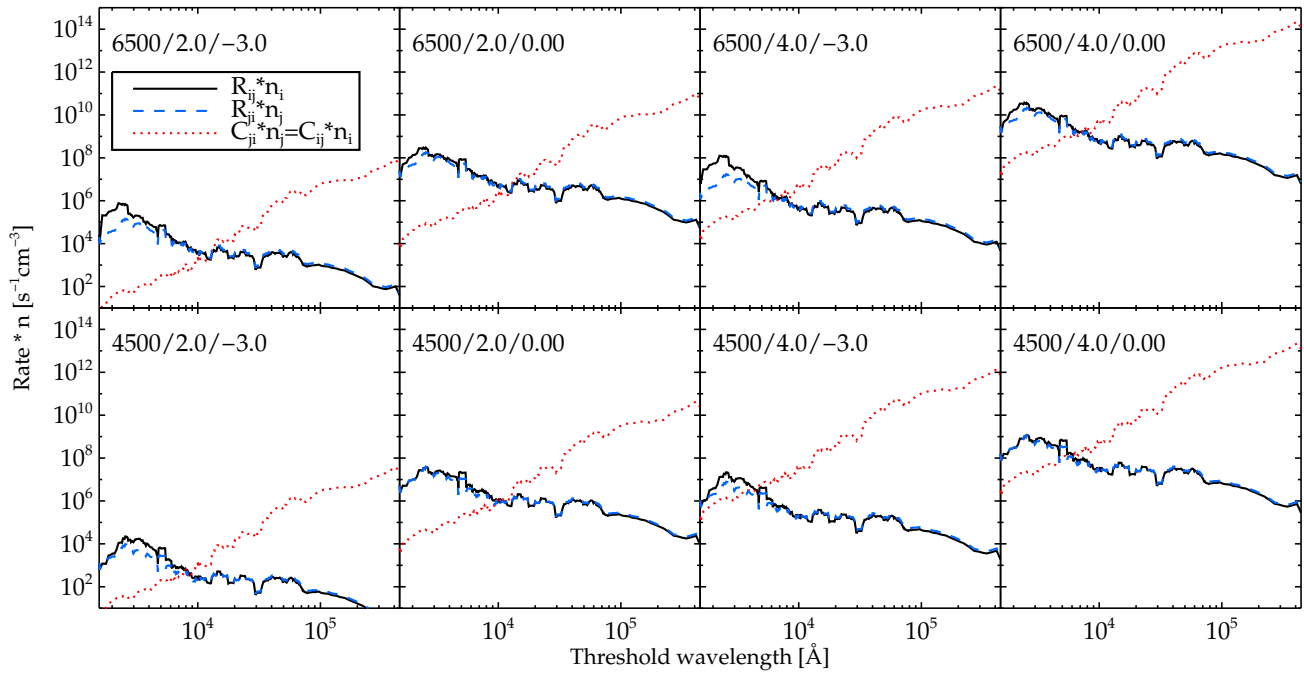}
\caption[]{The panels show the total number of Fe\,I--Fe\,II ionisation and recombination events per unit time and volume for selected models at line forming layers ($\rm log(tau_{500}=-2)$). The abscissa marks the threshold wavelength, $n_i$ is the LTE level population of the lower level, and $n_j$ the LTE population of the upper level of the corresponding transition, i.e. the ground state of Fe\,II. The curves have been constructed by applying a smoothing factor of ten to the individual values, so that each point on the curve represents the average of ten adjacent rates.} 
\label{Rates} 
\end{center} 
\end{figure*}

\subsection{Dependence on stellar parameters}{\label{sec:dependence}}
As Fig. \ref{delta} shows, the departures from LTE have a smooth and simple
dependence on stellar parameters. In summary, NLTE abundance corrections
increase with decreasing metallicity, increasing effective temperature, and
decreasing surface gravity. Here we do not aim to disentangle all the
contributing factors in order to explain the detailed behaviour quantitatively, but rather we highlight some apparent circumstances that enable a qualitative
understanding of the trends. 

We focus here on the perturbed ionisation balance, established by over-ionisation in Fe\,I bound-free transitions compared to LTE. We take the term overionisation to imply that $n_iP_{ij}-n_jP_{ji}>0$, where $n_i$ is the level
population of a bound state of Fe\,I, and $n_j$ is the ground state of Fe\,II  \citep[notations follow][]{Rutten03}. $P_{ij}=R_{ij}+C_{ij}$, where $R_{ij}$ is the radiative rate from level $i$ to
$j$ and $C_{ij}$ the corresponding collisional rate. Hence, over-ionisation means that the ionisation-recombination balance for a transition is shifted towards more efficient ionisation, which tends to de-populate the lower level with respect to LTE. Analogously, over-excitation is adopted for a transition whose excitation-de-excitation balance is shifted towards more efficient excitation.
Fig.\,\ref{Rates} demonstrates how the total photo-ionisation
and photo-recombination rates per unit volume ($n_iR_{ij}$ and $n_jR_{ji}$) vary with stellar
parameters and the threshold wavelength of the transition. Also shown are the
total collisional rates, which in LTE are related by $n_iC_{ij}=n_jC_{ji}$. 

The photo-ionisation rate, $R_{ij}$, is governed by the photo-ionisation cross-sections and the radiation field. In practice, the second dominates the energy dependence of $R_{ij}$, which exhibits a sharp increase from UV to optical wavelengths and thereafter declines mildly towards infra-red wavelengths \footnote{Low flux levels in the far UV in late-type stars suppresses the photo-ionisation rates for low-excited levels. Rates for transitions that bridge smaller energy gaps grow increasingly larger as they feel the effect of the stronger radiation field in the optical regime.}. The Fe\,I level population is maximal for the ground state and decreases exponentially with excitation energy. The product $n_iR_{ij}$ thus behaves as shown in Fig.\,\ref{Rates}, peaking in the UV regime. The recombination rates per unit volume ($n_jR_{ji}$) follow a very similar behaviour but deviate from the photo-ionisation rates as the radiation field deviates from the Planck function. 

Collisional rates, $C_{ij}$, on the other hand, depend on the bound-free collisional cross-sections and the availability of impact electrons and atoms.
The impact species have kinetic energy spectra that peak at energies corresponding to infra-red wavelengths \footnote{At typical temperatures, $kT\sim0.5\rm\,eV$, which corresponds to a kinetic energy peak at $2\rm\mu m$.} . In addition, the bound-free collisional cross-sections for electrons and hydrogen atoms increase with decreasing energy gap. These two dependencies outweigh the drop in level population, and the total rate uniformly increases towards longer wavelengths. The relative behaviour of the plotted curves is tightly connected to the sizes of the departures from LTE, through the equations of statistical equilibrium. 

Of course, all levels are also coupled by bound-bound radiative rates that can drive the level populations from LTE. Lines, or parts of lines, which are formed close to continuum-forming layers ($\tau_{500}\approx1$) in the atmosphere react analogously to bound-free transitions to the non-Planckian radiation field. Excitation is thus favoured above de-excitation in the UV and optical in the inner atmosphere, more specifically in layers which are optically thick to the radiation in strong line cores. The accumulated effect of the imbalance in the bound-bound rates is here to enhance the under-population of relevant levels of neutral iron. This can be understood because the levels most affected by over-excitation are the lowest excited, while those most affected by over-ionisation are moderately excited ($2-4$\,eV, see Fig.\,\ref{Rates}). The level populations of FeI are thus re-distributed by line transitions in a way that enhances under-population of relevant FeI levels ($<5$\,eV). We note that in shallow atmospheric layers, where the cores of strong lines become optically thin, different NLTE effects set in that instead favour de-excitation.

\begin{figure*} 
\begin{center} 
\includegraphics[scale=0.85,viewport=0cm 0cm 22cm 19cm,clip]{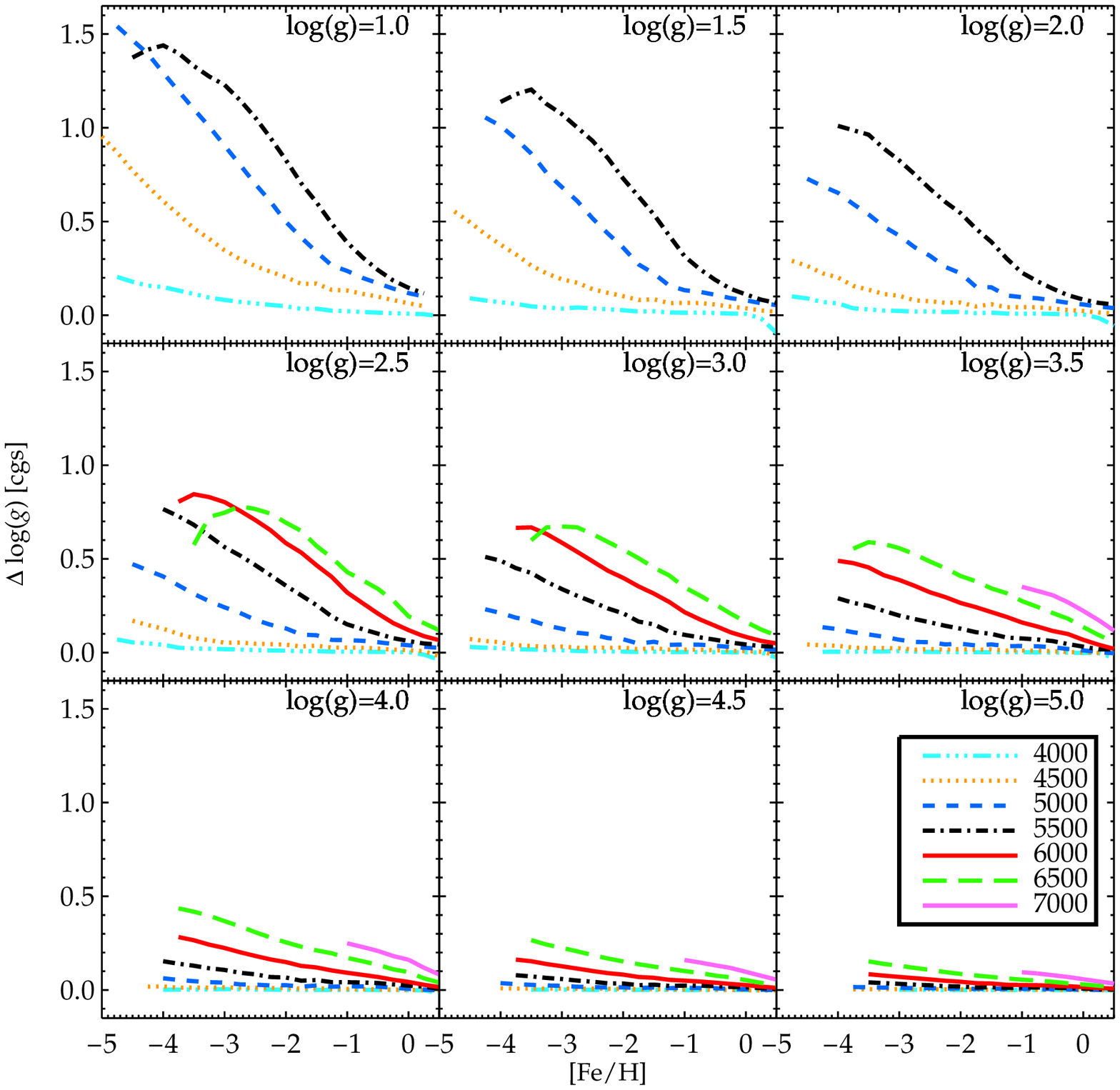}
\caption[]{The lines mark the estimated NLTE effect on surface gravities derived from ionisation balance of high-excitation, unsaturated Fe\,I and Fe\,II lines, given that other stellar parameters are fixed by independent means. All models shown have $\xi_{\rm t}=2.0$\,km/s.} 
\label{ion1} 
\end{center} 
\end{figure*}

We now proceed to interpret the NLTE dependencies on stellar parameters.
The over-ionisation is driven by the surplus of UV photons with respect
to the Planck function in the atmospheric radiation field. 
In turn, the size of the $J_\nu-B_\nu$-excess in the UV depends on
the steepness of the atmospheric temperature gradient and the efficiency of
metal line blocking in this part of the spectrum. We can therefore 
easily interpret the trend of increasing NLTE effects with decreasing metallicity.
It can be seen in Fig.\,\ref{Rates} that the gap between the photoionisation and 
photo-recombination curves grows larger at lower metallicity, while the relative 
contribution of radiative and collisional processes is approximately constant 
(the absolute rates are obviously lowered in accordance with the smaller 
amount of Fe atoms). 

The clear trend of increasing NLTE effects at lower surface gravity can
be understood from Fig.\,\ref{Rates} as due to the smaller relative strengths of 
collisional rates, scaling with the number density of iron atoms and the ionising impact species, 
with respect to radiative rates, that only scale with the number density of iron. 

The temperature gradient steepen sharply at higher effective temperatures, 
while the flux maximum moves towards shorter wavelengths. In Fig.\,\ref{Rates} we see
that higher effective temperatures again widen the gap between photoionisation and 
photo-recombination curves. In addition, the relative efficiency of collisions with 
respect to radiative transitions in lessened, since the increased amount of 
photons outnumber the increased amount and higher speed of the impact species. All in 
all, this explains why higher effective temperatures give rise to larger NLTE effects.

Finally, one might ask why only the strengths of Fe\,I lines are affected by NLTE and not 
Fe\,II lines, for which similar arguments as outlined above hold at least qualitatively. 
Bound-free transitions in Fe\,II are much less influential on the statistical 
equilibrium because of the large ionisation potential, hence low number density of Fe\,III. Furthermore, the three stellar-parameter 
dependencies (lower metallicity, lower surface gravity, and higher effective 
temperature) that give rise to increasing influence from the UV radiation field also act as to raise 
the relative fraction of singly ionised iron, which usually is the 
dominant species. A perturbation from the LTE populations clearly has a larger 
relative influence on a minority species. The well populated levels from which the visible Fe\,II transitions originate ($\la4\rm\,eV$) are generally close to thermalised throughout the whole atmosphere, while departures from LTE are apparent only above very high excitation energies ($\ga8$\,eV). 

\begin{figure*} 
\begin{center} 
\includegraphics[scale=0.85,viewport=0cm 0cm 22cm 19cm,clip]{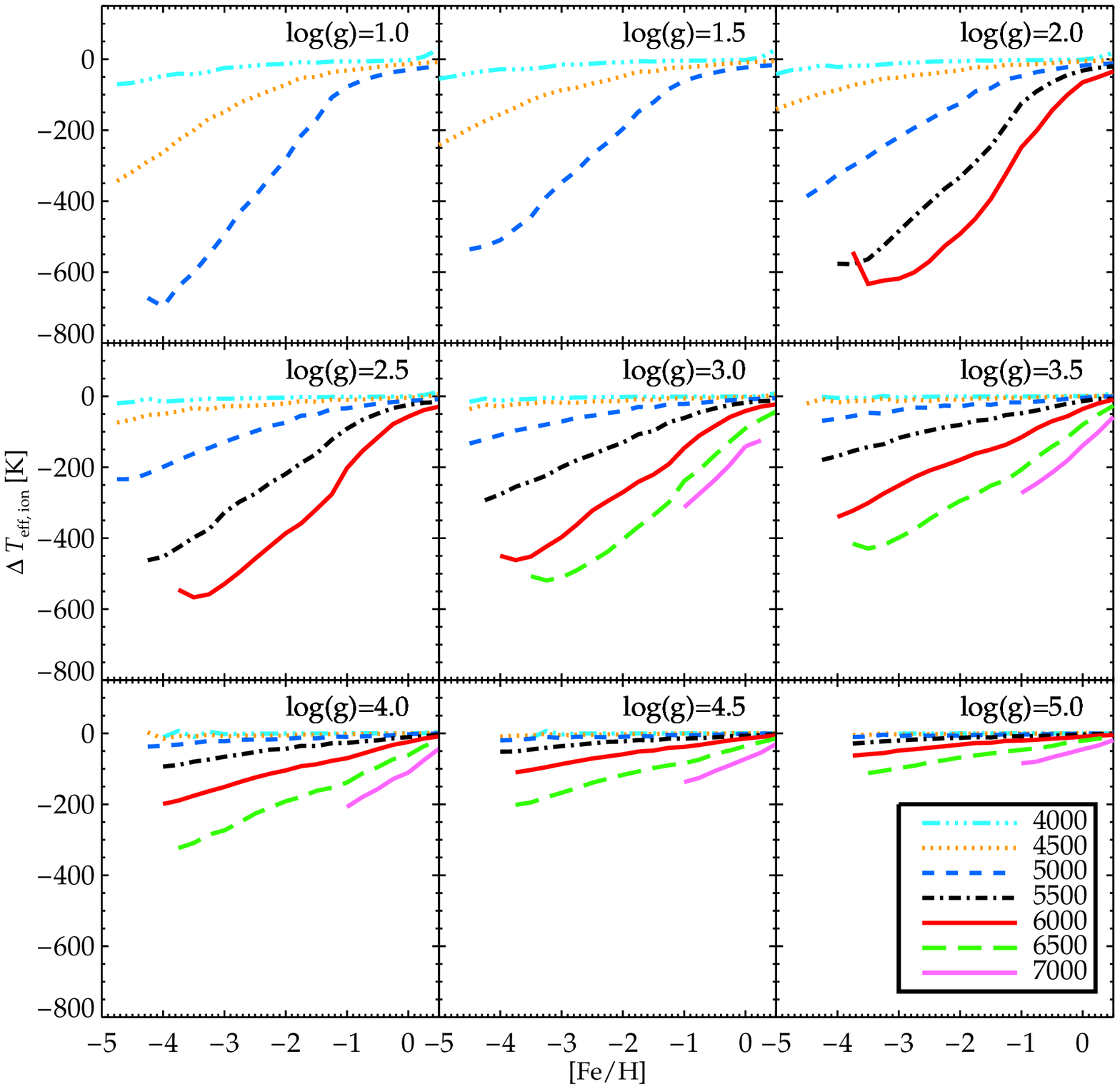}
\caption[]{The lines mark the estimated NLTE effect on effective temperatures derived from ionisation balance of high-excitation, unsaturated Fe\,I and Fe\,II lines, given that other stellar parameters are fixed by independent means. All models shown have $\xi_{\rm t}=2.0$\,km/s.} 
\label{ion2} 
\end{center} 
\end{figure*}

\begin{figure*} 
\begin{center} 
\includegraphics[scale=0.85,viewport=0cm 0cm 22cm 19cm,clip]{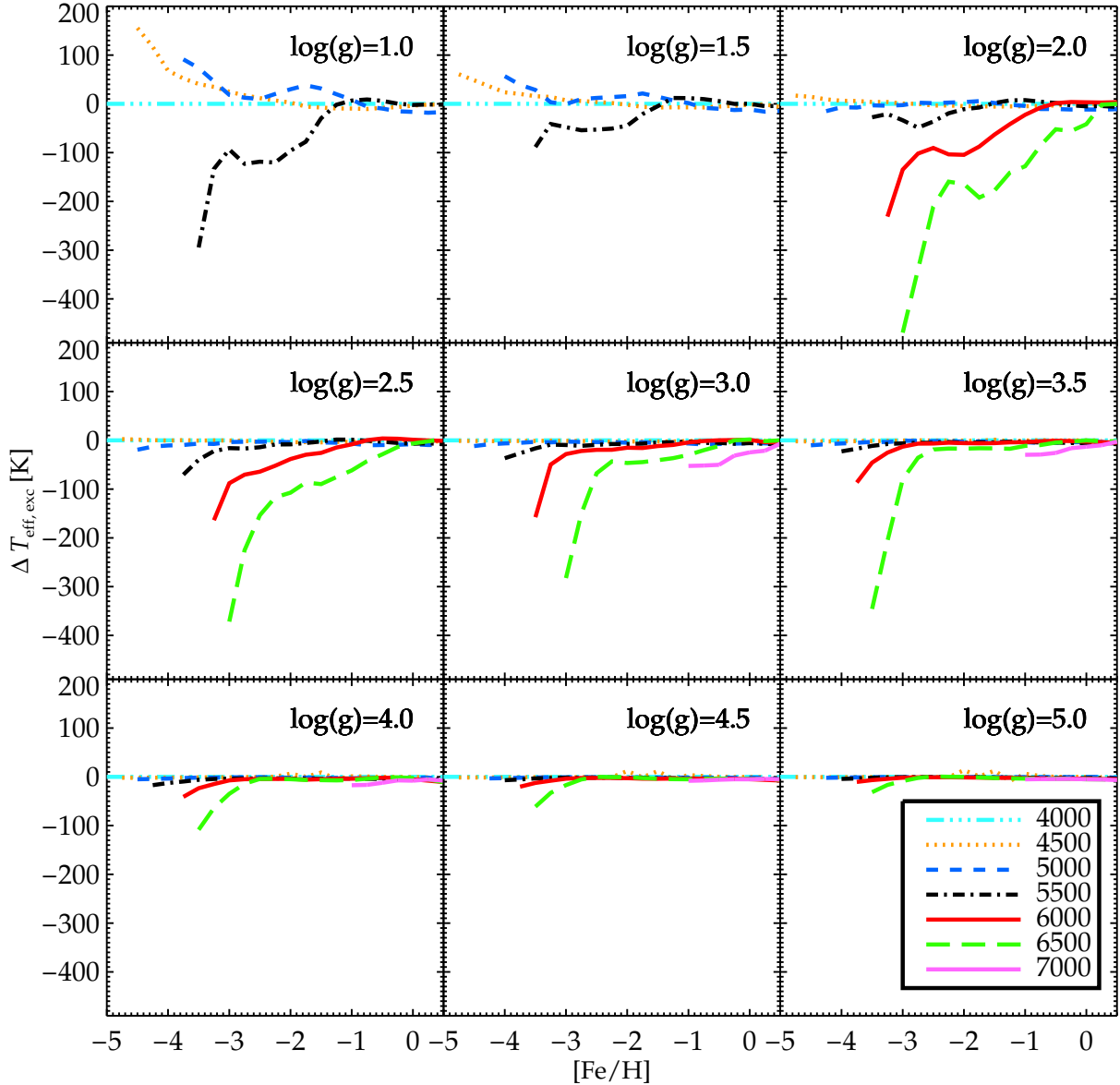}
\caption[]{The lines mark the estimated NLTE effect on effective temperatures derived from excitation balance of un-saturated Fe\,I lines, given that other stellar parameters are fixed by independent means. All models shown have $\xi_{\rm t}=2.0$\,km/s.} 
\label{exc} 
\end{center} 
\end{figure*}

\begin{figure*} 
\begin{center} 
\includegraphics[scale=0.85,viewport=0cm 0cm 22cm 19cm,clip]{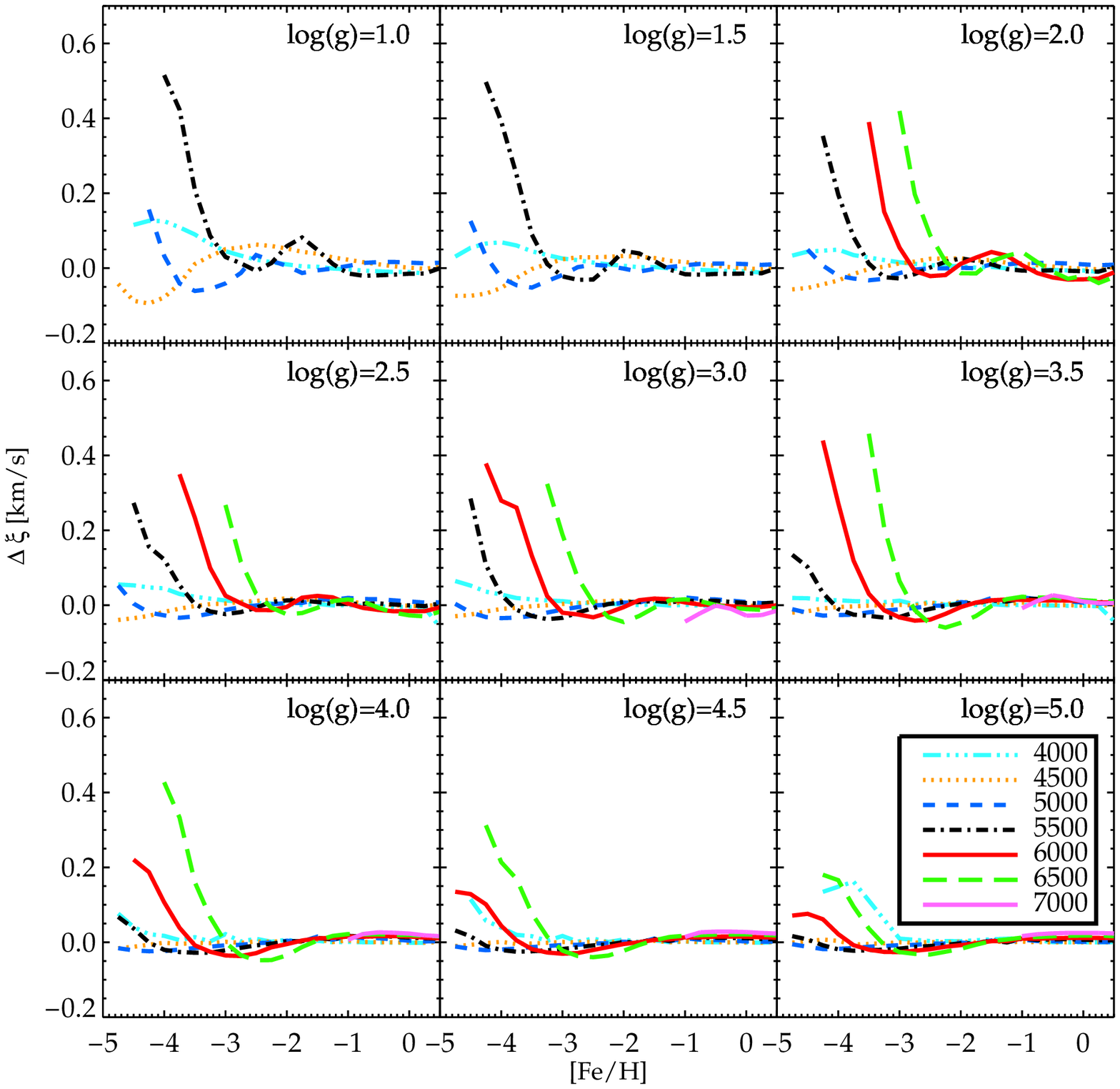}
\caption[]{The lines mark the estimated NLTE effect on microturbulence velocities derived from high-excitation Fe\,I lines, given that other stellar parameters are fixed by independent means. The reference value in LTE is $1\rm\,km\,s^{-1}$} 
\label{mic} 
\end{center} 
\end{figure*}

\subsection{Determining stellar
parameters}{\label{sec:determining}}

In the following subsection we will discuss the differences between
LTE and NLTE stellar parameters as inferred from Fe\,I and Fe\,II lines.

\subsubsection{Metallicity}
How NLTE affects the metallicity determination can be
straightforwardly estimated from Fig.\,\ref{delta}. Spectroscopic LTE analyses
underestimate the iron abundance derived from Fe\,I lines by anything between $0.0$ and $0.5$\,dex,
the exact value being strongly dependent on $T_{\rm eff}$, $\log g$ and [Fe/H]. 
Abundances derived from
Fe\,II lines are on average not affected at all in models with $\rm[Fe/H]>-3$. Even at
$\rm[Fe/H]=-4$, the corrections for the few lines that are still detectable are
smaller than 0.02\,dex. We conclude that LTE is a safe assumption for Fe\,II
lines in our parameter space, in line with several previous studies, e.g. \citet{Thevenin99,Korn03,Collet05,Mashonkina11a}, but see also \citet{Cram80}

\subsubsection{Surface gravity}
It is common to vary the model surface gravity until agreement is achieved 
between abundances derived from neutral and singly ionised iron lines, so called ionisation balance. Since Fe\,I lines are much less sensitive to surface gravity variations compared to Fe\,II lines, the metallicity is then practically based on the former. Since
Fe\,I lines are subject to significant NLTE effects, the LTE ionisation balance 
is not always realistic and the surface gravity as well as the metallicity will therefore be underestimated 
with this method.  We have assessed the
size of this effect by interpolating LTE curves-of-growth with varying surface
gravity onto the
NLTE equivalent widths found at each grid point. Since we are only interested in the difference between LTE and NLTE, the comparison can in principle be inverted and the LTE values chosen as the reference values. This would give very similar results, but since the NLTE equivalent widths are presumably more realistic it is more appropriate to use them as reference. For this comparison we again limit
ourselves to 
un-saturated, high-excitation FeI and FeII lines. 

The results, i.e. $\Delta \log g = \log g_{\rm NLTE}-\log g_{\rm LTE}$, are shown in Fig.\,\ref{ion1}. 
As a rule-of-thumb, 
the surface gravity is underestimated by a factor $x$ times the size of the NLTE 
correction for a given model, in logarithmic units. For dwarfs, $x\approx2.5$ while for giants the sensitivity increases slightly, $x\approx3$.   Thus for a metal-poor
turn-off star at $T_{\rm eff}=6500\,$K, $\rm log\it(g)\rm=4.0$, $\rm[Fe/H]=-2$, 
that has a typical NLTE abundance correction of 0.1\,dex, LTE analysis 
underestimates the surface gravity by 0.25\,dex. An important consequence of
this behaviour is that metal-poor 
dwarfs easily can be misclassified as subgiants. Similarly, a red giant star with $T_{\rm eff}=4500\,$K, $\rm log\it(g)\rm=1.5$, $\rm[Fe/H]=-5$ that has a NLTE abundance correction of 0.2\,dex would have its surface gravity underestimated by 0.6\,dex in LTE. 

\subsubsection{Effective temperature}
Here, we consider two different methods to infer the effective temperature, while the other parameters are kept fixed. First, given a reliable measurement of surface gravity \footnote{This can be provided by an accurate parallax measurement or from stellar evolution calculations. Both methods require a reasonable estimate on effective temperature, in which case the stellar parameter determination may need iteration.}, one may constrain effective temperature through ionisation balance. In this case, Fe\,I based abundance are effectively brought into agreement with the less temperature sensitive Fe\,II lines. The metallicity determination itself thus does not suffer heavy bias in the LTE case, but the optimised effective temperature will be overestimated, by the amount shown in Fig\,\ref{ion2}. The calculations were based on weak, high-excitation lines, as before.  

Secondly, the effective temperature can be inferred from the excitation balance of Fe\,I lines, flattening abundance trends with the excitation potential of the lower level of the transition (Fe\,II lines typically do not span a sufficient range in excitation potential in spectra of late-type stars).  The method exploits the larger temperature sensitivity of low excitation lines compared to high excitation lines, which simultaneously makes them more sensitive to departures from LTE. For 1D models, NLTE abundance corrections are nevertheless only mildly differential with excitation potential, but they become very significant for metal-poor \textless3D\textgreater\ models, as illustrated in Paper I. 

We have estimated the difference between 1D LTE and NLTE effective temperatures derived through excitation equilibrium, by again interpolating the LTE curves-of-growth onto NLTE results for all grid points, and enforced a flat trend with excitation potential. Again, we restrict ourselves to unsaturated lines, but the only restriction made in excitation potential is that at least one transition must originate from above $3.5$\,eV. Fig. \ref{exc} shows the typical difference in effective temperature derived in the two cases, i.e. $\Delta T_{\rm eff, exc}=T_{\rm eff, NLTE}-T_{\rm eff, LTE}$. Because NLTE abundance corrections are generally more positive for low excitation lines, the effective temperatures derived through LTE analysis are usually higher. Not surprisingly, the effect tends to increase in tandem with the NLTE effects, but the shape of curves are somewhat different.   Down to metallicities as low as $\rm[Fe/H]=-3$ the curves are rather flat and all types of dwarfs and the bulk of red giant branch stars are affected by less than 50\,K. Excitation temperatures derived with 1D models are thus not heavily influenced by NLTE effects except for horisontal branch stars and extremely metal-poor stars. However, we caution that even 1D NLTE results are not reliable in situations where 3D effects are pronounced, in particular if departures from LTE are minor.

\subsubsection{Microturbulence}
The NLTE abundance corrections show a dependence on line strength beyond saturation ($W_{\lambda}\ga50$\,m\AA\,), which will impact on the determination of microturbulence from Fe\,I lines. While the effect may be quite significant for individual lines, the typical change is very small. As a general rule, saturated ($50\la W_{\lambda}\la100$\,m\AA\,) lines have an abundance correction that is larger than for un-saturated lines, the difference being roughly factor of 5 less than the NLTE abundance correction itself in logarithmic units. Ultimately, this leads to lower  microturbulent velocities in LTE compared to NLTE. 

The size of this effect is illustrated in Fig.\,\ref{mic}. These results have been obtained analogously to the other parameters by calculating the microturbulence value that flattens the LTE abundance dependence on reduced equivalent width, using the NLTE equivalent widths as reference in each grid point. The calculations are restricted to high-excitation lines, to minimise a systematic bias as explained above. We also introduce an upper cut in line strength at $\log(W_\lambda/\lambda)=-4.6$ to exclude the damping part of the curve-of-growth. Similarly to the excitation balance, the effect is essentially negligible for dwarfs and giant stars above $\rm[Fe/H]\ge-3$. Extremely metal-poor stars are more affected, but generally they have few lines that are sensitive to this parameter.  

\subsection{Modelling uncertainties}

\subsubsection{Model atmospheres} 
Accurate absolute abundances in LTE and NLTE require realistic model atmospheres, hence 
spectrum synthesis based on 3D hydrodynamical models are expected to give superior results. 
Our first attempts to investigate this problem in LTE and NLTE using the averages of the 3D simulations of stellar convection were illustrated in Paper I for selected well-observed stars. We found that  the average metallicities derived from \textless3D\textgreater\ models are not dramatically different from 1D models, in particular in NLTE ($\la0.04$\,dex). Most sensitive to the model structure are abundances derived from saturated, low excitation lines in metal-poor stars. The greater similarity between NLTE abundances of 1D and \textless3D\textgreater\ models can be understood simply via the close resemblance of the mean radiation field throughout the atmospheres, as it is largely set by conditions at continuum optical depth $\tau_{\rm500\,nm}\ga1$, contrary to the Planck function that behaves very differently in shallow atmospheric layers. Paper I also illustrates how stellar parameters inferred through ionisation equilibrium of Fe\,I and Fe\,II lines agree well with NLTE modelling in 1D and  \textless3D\textgreater. We emphasize, however, that the excitation equilibrium established with \textless3D\textgreater\ models is generally superior to that of 1D models. Finally, it is worth to note that the uncovered dependencies of the departures from LTE on the stellar parameters 
are likely to hold true for any type of model atmosphere grid.  

\subsubsection{Atomic data}{\label{sec:error}} 

Our analysis focuses on systematic differences between LTE and
NLTE, which are not affected by the choice of oscillator strengths and
broadening data adopted for spectrum synthesis. According to our assessment, atomic data for level energies, radiative transition probabilities, and photo-ionisation cross-sections are sufficiently advanced and complete not to introduce large uncertainties in the modelling. Also, according to our tests the remaining uncertainties in the electron impact collision rates do not to introduce large uncertainties in the NLTE modelling. As mentioned in the introduction the most critical parameter is the efficiency of collisions with neutral hydrogen, which we constrained empirically in Paper I. Lowering the scaling factor $\rm S_H$ by one order of magnitude, from $1.0$ to $0.1$, increases the NLTE effects by typically a factor $2-4$, depending on the stellar parameters. The qualitative behaviour remains the same, but the influence of the radiative-rate imbalance that causes over-ionisation increases.  Paper I demonstrates that such large NLTE effects are unrealistic for metal-poor dwarfs and subgiant stars, but are potentially more successful for the metal-poor giant HD122563. Here, we have chosen to calculate the NLTE grid for $S\rm_H=1.0$, the most supported value, only. 

\subsubsection{The trace element assumption}{\label{sec:trace}} 
As mentioned in the introduction we solve the restricted NLTE problem, which
neglects all potential feedback that the perturbed Fe level
populations have on the atmospheric structure. This simplification is necessary
to make the computations tractable and the consequences have not been rigorously
investigated. The most noteworthy implicit assumptions are that the demonstrated changes in
excitation and ionisation balance have no significant implications for the free
electron density and the UV bound-bound and bound-free opacity. Given the known size
of the LTE
departures over the late-type star grid, we now proceed to make simple tests of
the extent of such a feedback.   

The main opacity change would be caused by the lower amount of neutral iron
compared to LTE, which in practice can be mimicked with an atmosphere of lower
metallicity. Furthermore, NLTE populations of iron would provide more free
electrons in the atmosphere. The increase in electron number density can be
straightforwardly estimated by the decrease in neutral iron, since each
additional iron ion corresponds to one additional electron. It turns out that
this effect is negligible over the whole stellar grid, since the free electron
density is typically much larger than that of neutral iron (hydrogen and metals with lower ionisation potential are the main electron donors). In the most extreme
case, the electron density of a cool, metal-rich giant would change by $\le1\%$, which is
insignificant in this context. We therefore disregard the second aspect and
focus our test on the opacity change.

We selected a few models from the grid and performed new NLTE calculations on
models with the same stellar parameters, but with a metallicity lower by the mean NLTE correction.  The lower metallicity was also adopted for the calculation of continuous and line background opacity. Thereafter we compare the LTE and NLTE
equivalent widths for a given abundance, obtained from the two different model
atmospheres. We perform the test on a metal-rich and metal-poor dwarf and giant with the same parameters as shown in Fig.\,\ref{Rates}. 

The solar metallicity models have a higher sensitivity to the metal content and are more affected by the mentioned inconsistency, even though the predicted sizes of the NLTE corrections are much smaller than for metal-poor stars. In particular, continuous opacities in the UV decrease when the model metallicity is lowered. We thus expect equivalent widths of Fe lines to increase due to lowered continuous opacities, while the NLTE effects are expected to increase and weaken the Fe\,I lines with respect to LTE. In practice, the NLTE equivalent widths inferred from the model with $T_{\rm eff}=6500\,$K and $\rm log\it(g)\rm=4.0$ increase by up to $1\%$ when the model metallicity (not the abundance used for computing synthetic line profiles) is lowered by 0.03\,dex, corresponding to the mean NLTE abundance correction. This is certainly a second-order effect, approximately ten times smaller than the NLTE effect itself. 

Note that our test implicitly assumes that all the ionisation balance of all other elements are similarly affected by NLTE as Fe, which is not the case. Only consistent NLTE model atmosphere calculations will tell for sure how realistic the trace element assumption is for Fe and other elements. \citet{Short05} present such a model for the Sun, displaying a dramatic increase in UV flux in NLTE compared to LTE, and a significantly more shallow atmospheric temperature structure at continuum forming regions. However, their comparison to observed absolute solar fluxes suggests that the opacity loss due to over-ionisation of iron-peak elements is over-estimated, or, alternatively, that other important sources of opacity are yet missing. 

\subsection{Comparison with other studies}{\label{sec:other}}

 \citet{Mashonkina11a} demonstrated in detail how recent developments in atomic data calculations have improved the NLTE modelling of iron lines. In particular, R.\,L.\,Kurucz's calculations of high excitation energy levels beyond reach of experiments have enabled a realistic coupling to the next ionisation state, without having to enforce an arbitrary thermalisation above a certain energy threshold \citep[see also][]{Gehren01}. As described in Paper I, our NLTE calculations for standard stars are in good quantitative agreement with those of \citet{Mashonkina11a}. In \citet{Mashonkina11b} the author presents a small grid of calculations for A-F type dwarfs of solar metallicity. Their
NLTE abundance corrections are of similar order of magnitude to ours, but differences for individual Fe\,I lines can amount to more than 0.05\,dex, which is comparable to the size of corrections for solar metallicity stars. There could be many reasons for this, among them different choices of scaling factor of hydrogen collisions. However, the dependence on surface gravity and effective temperature also show contradictory behaviour, which is surprising. While our NLTE corrections for e.g. Fe\,I 528.179\,nm  and 521.792\,nm increase gradually towards higher effective temperatures, those of \citeauthor{Mashonkina11b} rather display a flat behaviour or even decrease. Part of the solution might lie in the adoption of different oscillator strengths, placing the lines on slightly different parts of the curve-of-growth. 
 
Of earlier studies we highlight \citet{Thevenin99} who derived metallicities in NLTE for a size-able sample of metal-poor and metal-rich dwarfs and subgiants. They report a strong tendency of increasing NLTE effects with decreasing metallicity, akin to our findings but of greater magnitude. The reason for the difference is likely due to different model atoms; their atom lacked e.g. the important high excitation levels mentioned above. 

\subsection{Conclusions}
The most important NLTE effect for the determination of spectroscopic stellar parameters of late-type stars is the perturbed ionisation equilibrium of Fe\,I and Fe\,II lines. Departures from NLTE are significant for the derivation of Fe abundances from neutral lines, at the level of $\la$0.1\,dex in solar-metallicity stars, and $\la0.5$\,dex in metal-poor stars, while mostly insignificant for Fe\,II lines. Surface gravities obtained through the ionisation balance of neutral and singly ionised lines are consequently affected by $2.5-3$ times the typical NLTE abundance correction, in logarithmic units. Cooler stars are less affected than hotter, and dwarfs are less affected than giants, in a well-behaved pattern that can be understood by the variation of radiative and collisional rates over the HR diagram. 

Due to the demonstrated impact of \textless3D\textgreater\ effects on the Fe line formation \citep{Bergemann12}, we advice particular caution with temperatures derived through excitation equilibrium of metal-poor stars. 1D LTE effective temperatures are likely to be underestimated in this regime and 1D NLTE effective temperature possibly even more so. However, except for extremely metal-poor stars and high-temperature giants, the NLTE effects on the excitation balance in 1D are typically $\la50$\,K. Microturbulence values determined from Fe\,I lines are likewise very little affected by NLTE for these stars. 

Our grid calculations can be used to infer individual NLTE abundances for thousands of lines in late-type stellar spectra. It may be readily predicted that the field of Galactic archaeology subsequently will be confronted with a minor compression of metallicity distribution functions, due to the strictly increasing NLTE effects at lower metallicity. For example, a distribution function of the Galactic halo made up of $T_{\rm eff}=5000$\,K/$\log g=2.0$ giants would be affected by +$0.3$\,dex at the hyper metal-poor end, while the metal-rich end would remain essentially unchanged. On the other hand, the relative abundances of dwarf and giant stars may well be little affected, since dwarf samples are commonly hotter than giant samples, and consequently they are prone to departures from LTE of similar sizes. For example, the NLTE corrections of giants of the type just mentioned follow essentially the same behaviour as those of $T_{\rm eff}=6500$\,K/$\log g=4.0$ dwarfs.  

The propagated NLTE effects on Fe\,I lines on spectroscopic parameters are large enough to affect the classification of evolutionary states, ages, and masses. Forced with the requirement to meet ionisation balance of Fe\,I and Fe\,II, LTE analysis will underestimate surface gravities and/or overestimate effective temperatures, hence moving stars either upwards or leftwards in the HR diagram. Finally, trends of element ratios, [X/Fe], with respect to [Fe/H] are likely to undergo a change of shape. Assuming no change in element X, negative slopes will become flatter and positive slopes steeper. 

In the end, accurate Fe abundances is in general a pre-requisite for the accurate determination of abundances of all other elements.

\section*{Acknowledgments} We extend our thanks to M.\,Bautista for assisting with the implementation of the most recent data for photo-ionisation. 

%

\def\sun{\hbox{$\odot$}}
\def\la{\mathrel{\mathchoice {\vcenter{\offinterlineskip\halign{\hfil
$\displaystyle##$\hfil\cr<\cr\sim\cr}}}
{\vcenter{\offinterlineskip\halign{\hfil$\textstyle##$\hfil\cr
<\cr\sim\cr}}}
{\vcenter{\offinterlineskip\halign{\hfil$\scriptstyle##$\hfil\cr
<\cr\sim\cr}}}
{\vcenter{\offinterlineskip\halign{\hfil$\scriptscriptstyle##$\hfil\cr
<\cr\sim\cr}}}}}
\def\ga{\mathrel{\mathchoice {\vcenter{\offinterlineskip\halign{\hfil
$\displaystyle##$\hfil\cr>\cr\sim\cr}}}
{\vcenter{\offinterlineskip\halign{\hfil$\textstyle##$\hfil\cr
>\cr\sim\cr}}}
{\vcenter{\offinterlineskip\halign{\hfil$\scriptstyle##$\hfil\cr
>\cr\sim\cr}}}
{\vcenter{\offinterlineskip\halign{\hfil$\scriptscriptstyle##$\hfil\cr
>\cr\sim\cr}}}}}
\def\degr{\hbox{$^\circ$}}
\def\arcmin{\hbox{$^\prime$}}
\def\arcsec{\hbox{$^{\prime\prime}$}}
\def\utw{\smash{\rlap{\lower5pt\hbox{$\sim$}}}}
\def\udtw{\smash{\rlap{\lower6pt\hbox{$\approx$}}}}
\def\fd{\hbox{$.\!\!^{\rm d}$}}
\def\fh{\hbox{$.\!\!^{\rm h}$}}
\def\fm{\hbox{$.\!\!^{\rm m}$}}
\def\fs{\hbox{$.\!\!^{\rm s}$}}
\def\fdg{\hbox{$.\!\!^\circ$}}
\def\farcm{\hbox{$.\mkern-4mu^\prime$}}
\def\farcs{\hbox{$.\!\!^{\prime\prime}$}}
\def\fp{\hbox{$.\!\!^{\scriptscriptstyle\rm p}$}}
\def\cor{\mathrel{\mathchoice {\hbox{$\widehat=$}}{\hbox{$\widehat=$}}
{\hbox{$\scriptstyle\hat=$}}
{\hbox{$\scriptscriptstyle\hat=$}}}}
\def\sol{\mathrel{\mathchoice {\vcenter{\offinterlineskip\halign{\hfil
$\displaystyle##$\hfil\cr\sim\cr<\cr}}}
{\vcenter{\offinterlineskip\halign{\hfil$\textstyle##$\hfil\cr\sim\cr
<\cr}}}
{\vcenter{\offinterlineskip\halign{\hfil$\scriptstyle##$\hfil\cr\sim\cr
<\cr}}}
{\vcenter{\offinterlineskip\halign{\hfil$\scriptscriptstyle##$\hfil\cr
\sim\cr<\cr}}}}}
\def\sog{\mathrel{\mathchoice {\vcenter{\offinterlineskip\halign{\hfil
$\displaystyle##$\hfil\cr\sim\cr>\cr}}}
{\vcenter{\offinterlineskip\halign{\hfil$\textstyle##$\hfil\cr\sim\cr
>\cr}}}
{\vcenter{\offinterlineskip\halign{\hfil$\scriptstyle##$\hfil\cr
\sim\cr>\cr}}}
{\vcenter{\offinterlineskip\halign{\hfil$\scriptscriptstyle##$\hfil\cr
\sim\cr>\cr}}}}}
\def\lse{\mathrel{\mathchoice {\vcenter{\offinterlineskip\halign{\hfil
$\displaystyle##$\hfil\cr<\cr\simeq\cr}}}
{\vcenter{\offinterlineskip\halign{\hfil$\textstyle##$\hfil\cr
<\cr\simeq\cr}}}
{\vcenter{\offinterlineskip\halign{\hfil$\scriptstyle##$\hfil\cr
<\cr\simeq\cr}}}
{\vcenter{\offinterlineskip\halign{\hfil$\scriptscriptstyle##$\hfil\cr
<\cr\simeq\cr}}}}}
\def\gse{\mathrel{\mathchoice {\vcenter{\offinterlineskip\halign{\hfil
$\displaystyle##$\hfil\cr>\cr\simeq\cr}}}
{\vcenter{\offinterlineskip\halign{\hfil$\textstyle##$\hfil\cr
>\cr\simeq\cr}}}
{\vcenter{\offinterlineskip\halign{\hfil$\scriptstyle##$\hfil\cr
>\cr\simeq\cr}}}
{\vcenter{\offinterlineskip\halign{\hfil$\scriptscriptstyle##$\hfil\cr
>\cr\simeq\cr}}}}}
\def\grole{\mathrel{\mathchoice {\vcenter{\offinterlineskip\halign{\hfil
$\displaystyle##$\hfil\cr>\cr\noalign{\vskip-1.5pt}<\cr}}}
{\vcenter{\offinterlineskip\halign{\hfil$\textstyle##$\hfil\cr
>\cr\noalign{\vskip-1.5pt}<\cr}}}
{\vcenter{\offinterlineskip\halign{\hfil$\scriptstyle##$\hfil\cr
>\cr\noalign{\vskip-1pt}<\cr}}}
{\vcenter{\offinterlineskip\halign{\hfil$\scriptscriptstyle##$\hfil\cr
>\cr\noalign{\vskip-0.5pt}<\cr}}}}}
\def\leogr{\mathrel{\mathchoice {\vcenter{\offinterlineskip\halign{\hfil
$\displaystyle##$\hfil\cr<\cr\noalign{\vskip-1.5pt}>\cr}}}
{\vcenter{\offinterlineskip\halign{\hfil$\textstyle##$\hfil\cr
<\cr\noalign{\vskip-1.5pt}>\cr}}}
{\vcenter{\offinterlineskip\halign{\hfil$\scriptstyle##$\hfil\cr
<\cr\noalign{\vskip-1pt}>\cr}}}
{\vcenter{\offinterlineskip\halign{\hfil$\scriptscriptstyle##$\hfil\cr
<\cr\noalign{\vskip-0.5pt}>\cr}}}}}
\def\loa{\mathrel{\mathchoice {\vcenter{\offinterlineskip\halign{\hfil
$\displaystyle##$\hfil\cr<\cr\approx\cr}}}
{\vcenter{\offinterlineskip\halign{\hfil$\textstyle##$\hfil\cr
<\cr\approx\cr}}}
{\vcenter{\offinterlineskip\halign{\hfil$\scriptstyle##$\hfil\cr
<\cr\approx\cr}}}
{\vcenter{\offinterlineskip\halign{\hfil$\scriptscriptstyle##$\hfil\cr
<\cr\approx\cr}}}}}
\def\goa{\mathrel{\mathchoice {\vcenter{\offinterlineskip\halign{\hfil
$\displaystyle##$\hfil\cr>\cr\approx\cr}}}
{\vcenter{\offinterlineskip\halign{\hfil$\textstyle##$\hfil\cr
>\cr\approx\cr}}}
{\vcenter{\offinterlineskip\halign{\hfil$\scriptstyle##$\hfil\cr
>\cr\approx\cr}}}
{\vcenter{\offinterlineskip\halign{\hfil$\scriptscriptstyle##$\hfil\cr
>\cr\approx\cr}}}}}
\def\diameter{{\ifmmode\mathchoice
{\ooalign{\hfil\hbox{$\displaystyle/$}\hfil\crcr
{\hbox{$\displaystyle\mathchar"20D$}}}}
{\ooalign{\hfil\hbox{$\textstyle/$}\hfil\crcr
{\hbox{$\textstyle\mathchar"20D$}}}}
{\ooalign{\hfil\hbox{$\scriptstyle/$}\hfil\crcr
{\hbox{$\scriptstyle\mathchar"20D$}}}}
{\ooalign{\hfil\hbox{$\scriptscriptstyle/$}\hfil\crcr
{\hbox{$\scriptscriptstyle\mathchar"20D$}}}}
\else{\ooalign{\hfil/\hfil\crcr\mathhexbox20D}}%
\fi}}

\def\getsto{\mathrel{\mathchoice {\vcenter{\offinterlineskip
\halign{\hfil
$\displaystyle##$\hfil\cr\gets\cr\to\cr}}}
{\vcenter{\offinterlineskip\halign{\hfil$\textstyle##$\hfil\cr\gets
\cr\to\cr}}}
{\vcenter{\offinterlineskip\halign{\hfil$\scriptstyle##$\hfil\cr\gets
\cr\to\cr}}}
{\vcenter{\offinterlineskip\halign{\hfil$\scriptscriptstyle##$\hfil\cr
\gets\cr\to\cr}}}}}
\def\lid{\mathrel{\mathchoice {\vcenter{\offinterlineskip\halign{\hfil
$\displaystyle##$\hfil\cr<\cr\noalign{\vskip1.2pt}=\cr}}}
{\vcenter{\offinterlineskip\halign{\hfil$\textstyle##$\hfil\cr<\cr
\noalign{\vskip1.2pt}=\cr}}}
{\vcenter{\offinterlineskip\halign{\hfil$\scriptstyle##$\hfil\cr<\cr
\noalign{\vskip1pt}=\cr}}}
{\vcenter{\offinterlineskip\halign{\hfil$\scriptscriptstyle##$\hfil\cr
<\cr
\noalign{\vskip0.9pt}=\cr}}}}}
\def\gid{\mathrel{\mathchoice {\vcenter{\offinterlineskip\halign{\hfil
$\displaystyle##$\hfil\cr>\cr\noalign{\vskip1.2pt}=\cr}}}
{\vcenter{\offinterlineskip\halign{\hfil$\textstyle##$\hfil\cr>\cr
\noalign{\vskip1.2pt}=\cr}}}
{\vcenter{\offinterlineskip\halign{\hfil$\scriptstyle##$\hfil\cr>\cr
\noalign{\vskip1pt}=\cr}}}
{\vcenter{\offinterlineskip\halign{\hfil$\scriptscriptstyle##$\hfil\cr
>\cr
\noalign{\vskip0.9pt}=\cr}}}}}
\def\bbbr{{\rm I\!R}} 
\def\bbbm{{\rm I\!M}}
\def\bbbn{{\rm I\!N}} 
\def\bbbf{{\rm I\!F}}
\def\bbbh{{\rm I\!H}}
\def\bbbk{{\rm I\!K}}
\def\bbbp{{\rm I\!P}}
\def\bbbone{{\mathchoice {\rm 1\mskip-4mu l} {\rm 1\mskip-4mu l}
{\rm 1\mskip-4.5mu l} {\rm 1\mskip-5mu l}}}
\def\bbbc{{\mathchoice {\setbox0=\hbox{$\displaystyle\rm C$}\hbox{\hbox
to0pt{\kern0.4\wd0\vrule height0.9\ht0\hss}\box0}}
{\setbox0=\hbox{$\textstyle\rm C$}\hbox{\hbox
to0pt{\kern0.4\wd0\vrule height0.9\ht0\hss}\box0}}
{\setbox0=\hbox{$\scriptstyle\rm C$}\hbox{\hbox
to0pt{\kern0.4\wd0\vrule height0.9\ht0\hss}\box0}}
{\setbox0=\hbox{$\scriptscriptstyle\rm C$}\hbox{\hbox
to0pt{\kern0.4\wd0\vrule height0.9\ht0\hss}\box0}}}}
\def\bbbq{{\mathchoice {\setbox0=\hbox{$\displaystyle\rm
Q$}\hbox{\raise
0.05\ht0\hbox to0pt{\kern0.4\wd0\vrule height0.9\ht0\hss}\box0}}
{\setbox0=\hbox{$\textstyle\rm Q$}\hbox{\raise
0.05\ht0\hbox to0pt{\kern0.4\wd0\vrule height0.9\ht0\hss}\box0}}
{\setbox0=\hbox{$\scriptstyle\rm Q$}\hbox{\raise
0.05\ht0\hbox to0pt{\kern0.4\wd0\vrule height0.8\ht0\hss}\box0}}
{\setbox0=\hbox{$\scriptscriptstyle\rm Q$}\hbox{\raise
0.05\ht0\hbox to0pt{\kern0.4\wd0\vrule height0.8\ht0\hss}\box0}}}}
\def\bbbt{{\mathchoice {\setbox0=\hbox{$\displaystyle\rm
T$}\hbox{\hbox to0pt{\kern0.25\wd0\vrule height0.95\ht0\hss}\box0}}
{\setbox0=\hbox{$\textstyle\rm T$}\hbox{\hbox
to0pt{\kern0.25\wd0\vrule height0.95\ht0\hss}\box0}}
{\setbox0=\hbox{$\scriptstyle\rm T$}\hbox{\hbox
to0pt{\kern0.25\wd0\vrule height0.95\ht0\hss}\box0}}
{\setbox0=\hbox{$\scriptscriptstyle\rm T$}\hbox{\hbox
to0pt{\kern0.25\wd0\vrule height0.95\ht0\hss}\box0}}}}
\def\bbbs{{\mathchoice
{\setbox0=\hbox{$\displaystyle\rm S$}\hbox{\raise0.5\ht0\hbox
to0pt{\kern0.38\wd0\vrule height0.45\ht0\hss}\hbox
to0pt{\kern0.52\wd0\vrule height0.5\ht0\hss}\box0}}
{\setbox0=\hbox{$\textstyle \rm S$}\hbox{\raise0.5\ht0\hbox
to0pt{\kern0.38\wd0\vrule height0.45\ht0\hss}\hbox
to0pt{\kern0.52\wd0\vrule height0.5\ht0\hss}\box0}}
{\setbox0=\hbox{$\scriptstyle \rm S$}\hbox{\raise0.5\ht0\hbox
to0pt{\kern0.38\wd0\vrule height0.45\ht0\hss}\raise0.05\ht0\hbox
to0pt{\kern0.52\wd0\vrule height0.45\ht0\hss}\box0}}
{\setbox0=\hbox{$\scriptscriptstyle\rm S$}\hbox{\raise0.5\ht0\hbox
to0pt{\kern0.38\wd0\vrule height0.45\ht0\hss}\raise0.05\ht0\hbox
to0pt{\kern0.52\wd0\vrule height0.45\ht0\hss}\box0}}}}
\def\bbbz{{\mathchoice {\hbox{$\sf\textstyle Z\kern-0.4em Z$}}
{\hbox{$\sf\textstyle Z\kern-0.4em Z$}}
{\hbox{$\sf\scriptstyle Z\kern-0.3em Z$}}
{\hbox{$\sf\scriptscriptstyle Z\kern-0.2em Z$}}}}
\def\ts{\thinspace}
%
\def\aj{Astronomical Journal}
\def\actaa{Acta Astronomica}
\def\araa{Annual Review of Astronomy and Astrophysics}
\def\apj{Astrophysical Journal}
\def\apjl{Astrophysical Journal, Letters}
\def\apjs{Astrophysical Journal, Supplement}
\def\ao{Applied Optics}
\def\apss{Astrophysics and Space Science}
\def\aap{Astronomy and Astrophysics}
\def\aapr{Astronomy and Astrophysics Reviews}
\def\aaps{Astronomy and Astrophysics, Supplement}
\def\azh{Astronomicheskii Zhurnal}
\def\jcap{Journal of Cosmology and Astroparticle Physics}
\def\mnras{Monthly Notices of the Royal Astronomical Society}
\def\memsai{Memorie della Societiet\'a Astronomica Italia}
\def\memras{Memoirs of the Royal Astronomical Society}
\def\na{New Astronomy}
\def\nar{New Astronomy Review}
\def\pasa{Publications of the Astronomical Society of Australia}
\def\pra{Physical Review A: General Physics}
\def\prd{Physical Review D}%
\def\pasp{Publications of the Astronomical Society of the Pacific} 
\def\pasj{Publications of the Astronomical Society of Japan} 
\def\solphys{Solar Physics}
\def\ssr{Space Science Reviews}%
\def\zap{Zeitschrift f\"ur Astrophysik}
\def\nat{Nature}%
\def\aplett{Astrophysics Letters}
\def\jqsrt{Journal of Quantitiative Spectroscopy and Radiative Transfer}
\def\physrep{Physics Reports}
\def\physscr{Physica Scripta}
\let\astap=\aap
\let\apjlett=\apjl
\let\apjsupp=\apjs
\let\applopt=\ao

\bibliographystyle{mn2e} 

\bsp

\label{lastpage} \end{document}